\documentclass[twocolumn,showpacs,preprintnumbers,amsmath,amssymb]{revtex4}

\usepackage{graphicx}% Include figure files
\usepackage{dcolumn}% Align table columns on decimal point
\usepackage{bm}% bold math
\usepackage{amsmath}
\usepackage{url}

\begin{document}

\preprint{LA-UR 10-03289}

\title{Multi-Angle Simulation of Flavor Evolution in the Neutronization Neutrino Burst From\\an O-Ne-Mg Core-Collapse Supernova}% Working title

\author{John F. Cherry}
\affiliation{Department of Physics, University of California, San Diego, La Jolla, California 92093, USA}
\author{J. Carlson}
\affiliation{Theoretical Division, Los Alamos National Laboratory, Los Alamos, New Mexico 87545, USA}
\author{George M. Fuller}
\affiliation{Department of Physics, University of California, San Diego, La Jolla, California 92093, USA}
\author {Huaiyu Duan}
\affiliation{Theoretical Division, Los Alamos National Laboratory, Los Alamos, New Mexico 87545, USA}
\author{Yong-Zhong Qian}
\affiliation{School of Physics and Astronomy, University of Minnesota, Minneapolis, MN 55455}

\date{\today}

\begin{abstract}
We report results of the first $3\times3$ \lq \lq multi-angle\rq \rq \ simulation of the evolution of neutrino flavor in the core collapse supernova environment.  In particular, we follow neutrino flavor transformation in the neutronization neutrino burst of an O-Ne-Mg core collapse event.  Though in a qualitative sense our results are consistent with those obtained in $3\times3$ single-angle simulations, at least in terms of neutrino mass hierarchy dependence, performing multi-angle calculations is found to reduce the adiabaticity of flavor evolution in the normal neutrino mass hierarchy, resulting in lower swap energies.  Differences between single-angle and multi-angle results are largest for the normal neutrino mass hierarchy.  Our simulations also show that current uncertainties in the measured mass-squared and mixing angle parameters translate into uncertainties in neutrino swap energies.  Our results show that at low $\theta_{13}$ it may be difficult to resolve the neutrino mass hierarchy using the O-Ne-Mg neutronization neutrino burst.
\end{abstract}


\maketitle

\section{Introduction}
Here we present results of the first multi-angle, self-consistent, 3-flavor simulation of neutrino flavor evolution in the core collapse supernova environment. With an expected ultimate energy release of $ \sim0.1\,{\rm M}_{\odot}$ as neutrinos of all kinds, these supernovae have long been studied as potential sources of neutrinos which could provide probes of physics beyond the reach or scope of conventional terrestrial experiments.  Neutrino flavor oscillations are a case in point. Though we already know a great deal about neutrino mass-squared differences and flavor mixing properties from experiments, there are still fundamental neutrino mixing physics unknowns, {\it e.g.,} the neutrino mass hierarchy, mixing angle $\theta_{13}$, and CP violating phase $\delta$. Assessing how the known and unknown neutrino flavor oscillation physics affects the core collapse neutrino burst signature is a necessary step in understanding supernovae, as well as in gleaning insight into fundamental neutrino physics. 

We have chosen to examine the neutrino emission from supernovae originating from stars in the mass range $\sim8-10\,{\rm M}_{\odot}$.  Supernova progenitors in this mass range are expected to be relatively common, comprising $25\%$ or more of core collapse supernova events (taking the number of stars per unit mass to be $\propto m^{-2.35}$, using the Salpeter initial mass function).  These relatively light supernova progenitors develop O-Ne-Mg cores, which undergo gravitational collapse triggered by electron capture on Ne and Mg isotopes.  This collapse leaves a thin envelope above the proto-neutron star remnant, in contrast to the envelopes of more massive stars that have undergone the gravitational collapse of an Fe core.  

Progenitors of O-Ne-Mg core-collapse supernovae are characterized by a steep matter density gradient above the core, which may have shed most of the hydrogen envelope prior to core collapse due to pulsation and winds.  These characteristics lead to two salient features of O-Ne-Mg core-collapse events.  First, in contrast to the Fe core-collapse case, the bounce shock in models of these supernovae is not severely hindered by the pressure of material falling onto the core.  Consequently, explosion by direct neutrino driven shock re-heateing is obtained~\cite{Mayle88,Kitaura06,Dessart06}.  Second, the steep matter gradient above the core allows neutrino self-coupling (neutral-current neutrino-neutrino forward exchange scattering) to engineer collective neutrino flavor transformation~\cite{Fuller87,Notzold:1988fv,Pantaleone92,Fuller:1992eu,Qian93,Samuel:1993sf,Qian95,Kostelecky:1995rz,Samuel:1996rm,Pastor02,Pastor:2002zl,Sawyer:2005yg,Fogli07} during the early phase of neutrino emission, including shock break-out and the attendant neutronization neutrino burst~\cite{Duan08,Dasgupta:2008qy}.

However, there is another consequence of the steep matter density gradient.  A steep matter gradient necessitates accounting for the physics of the interference between neutrino flavor mixing scales. Essentially, Mikheyev-Smirnov-Wolfenstein (MSW) resonances can occur very near to each other in physical and energy space when the matter density profile drops off rapidly with radius.  The near overlapping of resonances for the solar and atmospheric vacuum neutrino mass-squared differences, $\Delta m^{2}_{\odot} = 7.6 \times 10^{-5}\,{\rm eV}^{2}$ and $\Delta m^{2}_{\rm atm} = 2.4\times 10^{-3}\,{\rm eV}^{2}$, respectively, dictates that a full $3\times3$ flavor treatment must be employed.  Because these resonances occur close to the core, and because in this case the number density of neutrinos emitted from the core will fall much more slowly than the local matter density, the anisotropic nature of the supernova environment necessitates a multi-angle simulation that includes trajectory-dependent neutrino self-coupling~\cite{Duan06a,Duan06b,Duan06c,Duan07a,Duan07b,Duan07c,Duan08,Duan:2008eb,Duan:2010bg}. 

Nonlinear neutrino flavor transformation in supernovae is a subject in its infancy and, by necessity, much of the discussion in the supernova community is focused on technical issues on how to calculate it.  To that end this paper raises an important point, that a \lq\lq standard\rq\rq\ simplifying treatment of the problem, the widely used \lq\lq single-angle\rq\rq\ approximation, may not reproduce the results of more sophisticated \lq\lq multi-angle\rq\rq\ calculations. 

Reference~\cite{Duan08}, which employs the single-angle approximation, found two flavor swaps in the normal mass hierarchy for the O-Ne-Mg core collapse neutronization burst, one for each mass-squared splitting, but only one flavor swap in the inverted mass hierarchy.  The general features found in the results of Ref.~\cite{Duan08} seem to agree with the semi-analytical analysis in Ref.~\cite{Dasgupta:2008qy}.  Reference~\cite{Dasgupta09} has pointed out that in single-angle calculations, under some circumstances, a single mass-squared splitting can give rise to two spectral swaps.  However, Ref.~\cite{Fried10} demonstrated recently that full $3\times3$ simulations, as opposed to a sequence of $2 \times 2$ level crossings, may be necessary to understand the formation or suppression of these multiple flavor swaps.  

As in Ref.~\cite{Duan08}, to better understand multi-angle effects in $3 \times 3$ neutrino flavor oscillation scenarios and to enable direct comparison between single-angle and multi-angle caculations, we have chosen to simulate the epoch of the neutronization neutrino burst.  This corresponds to an epoch only some $\sim10\,{\rm ms}$ after core bounce, when the shock wave propagates through the neutrino sphere.  The total neutrino luminosity at this epoch is comprised predominantly, but not exclusively, of electron type neutrinos $\left( \nu_{\rm e} \right)$ left over from core collapse-driven neutronization.

In section II we discuss issues associated with 3-neutrino flavor oscillations in vacuum and in medium.  In section III we discuss our particular numerical methodology for simulating multi-angle, 3-neutrino flavor development in supernovae.  In section IV we discuss results for the neutronization burst in O-Ne-Mg core collapse events.  Conclusions and prospects for future work are discussed in section V.

\section{supernova neutrino flavor transformation}

Neutrino oscillations in vacuum arise because the weak interaction (flavor) eigenstates for these particles $\vert \nu_{\alpha}\rangle$ are not coincident with their energy (mass) eigenstates $\vert \nu_{\rm i}\rangle$. Here $\alpha={\rm e},\mu ,\tau$ and $i = 1,2,3$ refer to the flavor states and mass eigenvalues $m_{\rm i}$, respectively. The neutrino mass-squared differences are $\Delta m^{2}_{\rm ij} = m^{2}_{\rm i} - m^{2}_{j}$. In vacuum the energy eigenstates are related to the flavor eigenstates by the Maki-Nakagawa-Sakata (MNS) matrix $U$: $\vert \nu_{\alpha}\rangle = \sum_{i}U^{*}_{\alpha \rm i}\vert\nu_{i}\rangle$;  where $U_{\alpha \rm i}$ are the elements of the unitary transformation matrix.  This transformation has four free parameters: three mixing angles and a $CP$-violating phase $\delta$.  Solar, atmospheric, reactor, and accelerator neutrino experiments have measured two of these parameters: $\theta_{12} \approx 0.59$, $\theta_{23} \approx \pi/4$.  The best current experimental constraints show that $\theta_{13} \leq 0.2$ ($2\sigma$ limit)~\cite{Ahluwalia97}.  The $CP$-violating phase $\delta$ remains unconstrained.  

The ordering (hierarchy) of the neutrino mass-squared differences also remains undetermined. There are two possible configurations for a set of $3$ mass states with two mass-squared splittings. The \lq \lq normal\rq \rq \ mass hierarchy has the solar neutrino mass-squared split below the atmospheric split.  In this case the vacuum mass states are ordered from lowest to highest as $m_{1}$, $m_{2}$, and $m_{3}$, with $\Delta m^{2}_{\odot} = m^{2}_{2} - m^{2}_{1}$ and $\Delta m^{2}_{\rm atm} = m^{2}_{3} - m^{2}_{2}$.  By contrast, the  \lq \lq inverted\rq \rq \ mass hierarchy is where the solar neutrino mass-squared split lies above the atmospheric split.  In this scheme the vacuum mass states are ordered from lowest to highest as $m_3$, $m_{1}$, and $m_{2}$, with $\Delta m^{2}_{\odot} = m^{2}_{2} - m^{2}_{1}$ and $\Delta m^{2}_{\rm atm} = m^{2}_{1} - m^{2}_{3}$.

%For a pair of neutrino states that are maximally mixed, or nearly so, as are neutrino vacuum mass states 2 and 3, we can construct a convenient basis which we refer to as the \lq \lq hybrid\rq\rq\ basis, or alternatively the \lq\lq interaction\rq\rq\ basis~\cite{Balantekin:2000hl,Caldwell:2000db}.  This hybrid basis is: $\vert\nu_{\rm e}\rangle$, $\vert\nu_{\rm x}\rangle = \left( \vert\nu_{\mu}\rangle + \vert\nu_{\tau}\rangle \right) / \sqrt{2}$, and $\vert\nu_{\rm y}\rangle = \left( \vert\nu_{\mu}\rangle - \vert\nu_{\tau}\rangle \right) / \sqrt{2}$.  Constructed in this fashion, the hybrid neutrino states $ \vert\nu_{\rm x}\rangle$ and $\vert\nu_{\rm y}\rangle$ correspond to the vacuum mass eigenstates $\vert\nu_{2}\rangle$ and $\vert\nu_{3}\rangle$, respectively.  Because $ \vert\nu_{\mu}\rangle$ and $\vert\nu_{\tau}\rangle$ neutrino flavor states experience virtually the same interactions in the supernova environment, this correspondence is maintained in medium, as well as in vacuum.

Neutrino oscillations in medium can differ significantly from the vacuum case. For the supernova environment above the proto-neutron star the neutrinos are nearly all free streaming. In this regime we therefore can neglect inelastic processes and associated de-coherence. Consequently we follow only coherent, elastic neutrino interactions. In this limit we can make the mean-field, coherent forward-scattering approximation, where neutrino flavor evolution is governed by a Schr\"odinger-like equation of motion. For example, we can represent the flavor state of neutrino $n$ by $\vert\psi_{\nu,n}\rangle$.  (The relationship between neutrino and antineutrino flavor states and the corresponding flavor isospin spinors is discussed in Ref.~\cite{Duan06c}.)  The evolution of this flavor state in the mean field coherent limit is then
\begin{equation}
i {{\partial \vert\psi_{\nu,n}\rangle  }\over{\partial t  }}=\hat{H} \vert\psi_{\nu,n}\rangle
\label{Schr}
\end{equation} 
%\begin{equation}
%i\frac{d}{dt}\vert\psi\rangle = H\vert\psi\rangle \text{,}
%\label{eqn:mfaprx}
%\end{equation}
where $t$ is an Affine parameter along neutrino $n$'s world line and $\hat{H}$ is the appropriate flavor-changing Hamiltonian along this trajectory: $\hat{H} = \hat{H}_{\rm vac} + \hat{H}_{\rm mat} + \hat{H}_{\nu\nu}$.  Here $\hat{H}_{\rm vac}$, $\hat{H}_{\rm mat} $, and $\hat{H}_{\nu\nu}$ are the vacuum, neutrino-electron/positron charged current forward exchange scattering, and neutrino-neutrino neutral current forward exchange scattering (neutrino self-coupling) contributions, respectively, to the overall Hamiltonian.

There are two different mass-squared difference scales as discussed above, the atmospheric $\Delta m^{2}_{\rm atm}$ and solar $\Delta m^{2}_{\odot}$ splittings.  In what follows we refer to neutrino mixing at a particular point as being on the \lq\lq $\Delta m^{2}_{\odot}$ scale\rq\rq\ or \lq\lq $\Delta m^{2}_{\rm atm}$ scale\rq\rq . By this we mean that the neutrino flavor transformation at this point is taking place mostly through $\nu_{\rm e} \rightleftharpoons \nu_{\mu ,\tau}$ mixing in that part of the unitary transformation corresponding to $\Delta m^{2}_{\odot}$ or $\Delta m^{2}_{\rm atm}$, respectively.  This terminology is a holdover from the standard adiabatic MSW case, where $\Delta m^{2}_{\odot}/2E_{\nu}$ and $\Delta m^{2}_{\rm atm}/2E_{\nu}$ essentially pick out density regions and neutrino energies $E_{\nu}$ where mixing is large (i.e., MSW resonance at $\Delta m^{2}/2E_{\nu} \approx \sqrt{2}G_{\rm F}n_{\rm e}$, where $n_{\rm e}$ is the net electron number density).

The coherent neutrino-neutrino forward exchange scattering Hamiltonian $\hat{H}_{\nu\nu}$ produces vexing nonlinear coupling of flavor histories for neutrinos on intersecting trajectories. This is the pivotal complication encountered when attempting to calculate the evolution of the supernova neutrino flavor field. Note that $\hat{H}_{\nu\nu}$ gives rise to both flavor-diagonal and off-diagonal potentials. In turn, each of these potentials is neutrino intersection angle-dependent, reflecting the $V-A$ structure of the underlying current-current weak interaction Hamiltonian in the low momentum transfer limit.  For neutrinos in state $l$ on a trajectory with unit tangent vector $\hat{k}_l$, the neutrino self-coupling Hamiltonian is given by a sum over neutrinos and antineutrinos $m$:
\begin{eqnarray}
& \hat{H}_{\nu\nu,l} & 
 =  \sqrt{2}\,G_{\rm F}\,\sum_m{\left( 1-\hat{k}_l\cdot\hat{k}_m \right) n_{\nu,m}\, \vert\psi_{\nu,m}\rangle\, \langle\psi_{\nu,m}\vert} 
\nonumber
\\
                 &  - &\sqrt{2}\,G_{\rm F}\, \sum_m{\left( 1-\hat{k}_l\cdot\hat{k}_m \right) n_{\bar\nu,m}\, \vert\psi_{\bar\nu,m}\rangle\, \langle\psi_{\bar\nu,m}\vert}
 \label{arrat}
\end{eqnarray}
where $\hat{k}_m$ is the unit trajectory tangent vector for neutrino or antineutrino $m$.  Here $n_{\nu,m}$ is the local number density of neutrinos in state $m$.  In the supernova environment, the effect of neutrino self-coupling may be to engender collective neutrino flavor transformation, the general features of which are reviewed in Ref.~\cite{Duan:2010bg}.  
 %\begin{widetext} 
%\begin{eqnarray}
%\vert\psi_{\nu_\alpha}(t)\rangle\langle\psi_{\nu_\alpha}(t)\vert   
%& = & {\vert a_{1\alpha}(t)\vert}^2\vert\nu_1(t)\rangle\langle\nu_1(t)\vert
%+{\vert a_{2\alpha}(t)\vert}^2\vert\nu_2(t)\rangle\langle \nu_2(t)\vert
%\nonumber
%\\
%& + &{a_{1\alpha}(t)a_{2\alpha}^\ast}(t)\vert\nu_1(t)\rangle\langle\nu_2(t)\vert 
%+{a_{1\alpha}^\ast(t)a_{2\alpha}(t)}\vert\nu_2(t)\rangle\langle \nu_1(t)\vert. 
%\label{singleop}
%\end{eqnarray}
%\end{widetext}

At a particular epoch, and with a specified matter density profile and specified initial neutrino fluxes and energy spectra, numerical calculation of the neutrino flavor field above the proto-neutron star would require a self-consistent solution of Eq.\ \ref{Schr} for neutrinos on all trajectories. This must include a prescription for treating $\hat{H}_{\nu\nu}$ in Eq.\ \ref{arrat}, which couples flavor evolution on intersecting neutrino trajectories. This can be done by adopting the so-called \lq\lq single-angle\rq\rq\ approximation~\cite{Qian95}, where the flavor evolution along a specified neutrino trajectory ({\it e.g.,} the radial trajectory) is taken to apply along all other trajectories at corresponding values of the Affine coordinate on those world lines. Alternatively, a \lq\lq multi-angle\rq\rq\ treatment can be employed where no approximations are used in the self-consistent evaluation of the Hamiltonian in Eq.\ \ref{arrat}. Though especially difficult to carry out computationally in a full 3-neutrino mixing scheme, some supernova scenarios may require such a treatment.  To this end, we have developed numerical techniques to implement this approach.

\section{Methodology}
The numerical codes used for simulating neutrino flavor evolution in the calculations reported in this paper are the FLAT code and the BULB code.  These codes, and related schemes to solve for the flavor evolution of core collapse supernova neutrinos, are discussed in Ref.~\cite{Duan:2008eb}.  

In order to parallelize the nonlinearly-coupled differential equations which describe neutrino flavor evolution, BULB employs a specific geometric representation of the region above the neutrino sphere.  In this representation all neutrinos are assumed to be emitted from a hard spherical shell, and propagate through a one dimensional, spherically symmetric distribution of matter.  Of course, spherical symmetry implies that no aspect of any coherent neutrino forward scattering potential depends on polar or azimuthal coordinate.  This high degree of symmetry allows the neutrino emission to be broken down and grouped by species, energy, and emission angle.  Here we define the emission angle, $\vartheta_{\rm R}$, to be the angle between the neutrino direction and the vector normal to the surface of the neutrino sphere at the neutrino emission point.  Our choice of parallelization is to assign all of the neutrino species and energies for a single emission angle bin to a single core (multiple angle bins can be assigned to a single core if desired, though this is less efficacious).  This allows for efficient, fine grained parallelization.  This feature is critical, as the emission angle dimension in the simulation may require high resolution in order to achieve convergence of the whole calculation.

To initialize the simulation, neutrinos are allocated to each energy-angle bin according to the species-specific luminosity and neutrino energy spectral type (blackbody or \lq \lq pinched\rq \rq ).  From there, BULB employs Heun's method, which is a second order predictive-corrective algorithm, to compute the flavor evolution of the neutrino states.  To minimize inter-node traffic, each process sums up the neutrino density matrix elements for the energy-angle bins it is responsible for, and sends this reduced data back to a central process.  This central process gathers all of the neutrino density matrix data and computes a unique forward scattering Hamiltonian $H_{\rm 0}$, including $H_{\rm vac}$, $H_{\rm mat}$, and $H_{\nu\nu}$, for each energy-angle bin.  These Hamiltonians are then redistributed to the appropriate processes so that each process can evaluate $\vert\psi\rangle_{\Delta t} = \exp\left(-iH_{\rm 0}\Delta t\right) \vert\psi\rangle_{\rm 0}$, where $\vert\psi\rangle_{\rm 0}$ is the initial wave function, and $\vert\psi\rangle_{\Delta t}$ is the resultant wave function after a single step in Affine parameter $\Delta t$ along a given neutrino trajectory.  Once that is completed, the central process again collects all of the neutrino energy-angle states $\vert\psi\rangle_{\Delta t}$ to compute a new set of Hamiltonians $H_{\Delta t}$ appropriate for the end point of the step.  These are then sent back out to the individual processes, which now use the average of $H_{\rm 0}$ and $H_{\Delta t}$ to compute the evolved neutrino state:
\begin{equation}
\vert\psi\rangle_{\rm final} = \frac{1}{2} \left[ \exp\left(-iH_{\rm 0}\Delta t\right) \\
+ \exp\left(-iH_{\Delta t}\Delta t\right) \right] \vert\psi\rangle_{\rm 0} \text{.}
\label{eqn:Psifinal}
\end{equation}

In order to check for convergence, a second round of computations are made using the Heun method to evolve the state $\vert\psi\rangle_{\rm 0}$, using the above algorithm twice, with a step size ${\Delta t}/{2}$.  Every process checks that both final states for $\vert\psi\rangle_{\rm final}$ agree to within a predefined error tolerance, usually chosen to be 1 part in $10^{8}$.  If the final states for single energy-angle bins do not converge, the step size is halved and the process is started over.  This guarantees that the entire set of neutrino flavor states are tracked consistently.  Periodically, $\Delta t$ is increased by a factor of 2 to make sure that the simulation is operating with an efficiently chosen time step.

Convergence of the overall calculation is checked by comparing results with different error tolerances and differing numbers of energy and angle bins.  Generally, large numbers of energy and angle bins, typically $\sim (1000\ \text{angle bins}) \times (400\ \text{energy bins})$, are required to achieve good results.  

We have validated our simulations by performing them with two numerically distinct codes, FLAT and BULB, and comparing our results~\cite{Duan:2008eb}.  Using the same set of initial conditions, both codes agree with each other at the level of $0.1\%$ when comparing the final neutrino flavor states.  FLAT works in analogous fashion to BULB, though these codes use different integration algorithms for handling the partial differential equations governing neutrino flavor evolution.  This procedure has enabled the development of a full $3\times3$ multi-angle treatment for supernova neutrino flavor transformation.

\section{Results}

\subsection{Overview}
\begin{figure*}
\centering
\includegraphics[scale=.85]{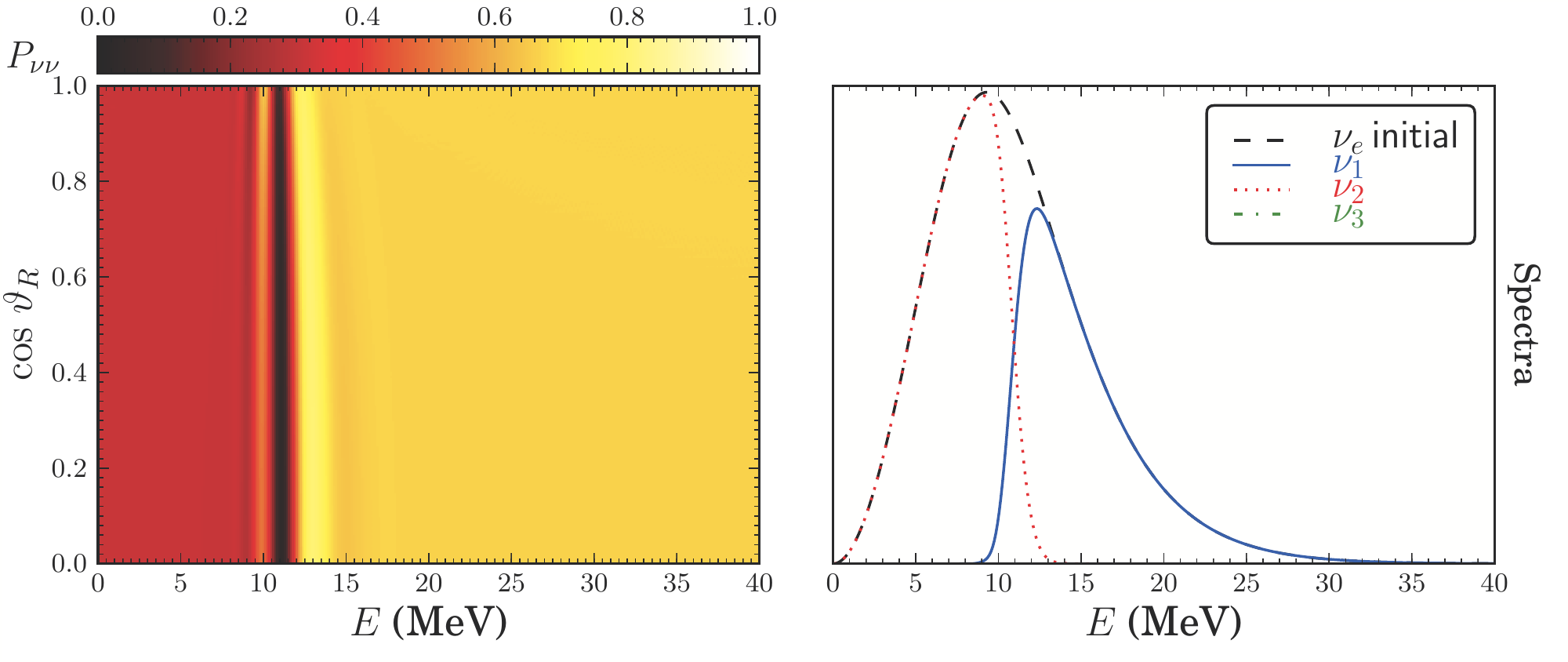}%scale = .63
\caption{Left panel: electron neutrino survival probability $P_{\nu_{\rm e}\nu_{\rm e}}$ (color/shading key at top left) for the inverted mass hierarchy is shown as a function of cosine of emission angle, $\cos{\vartheta_{\rm R}}$, and neutrino energy, $E$ in MeV, plotted at a radius of $r=5000\,{\rm km}$.  Right:  mass basis (key top right, inset) emission angle-averaged neutrino energy distribution functions versus neutrino energy, $E$.  The dashed curve gives the initial $\nu_{\rm e}$ emission angle-averaged energy spectrum.  Movies of this simulation can be found at the URL in Ref.~\cite{Duan:web}.  Each frame of the movie shows a representation of the neutrino survival probability in various different bases at a fixed radius above the core.  Each successive frame is $1\,{\rm km}$ further out from the initial radius of $r_{\rm init} = 900\,{\rm km}$ out to the final radius $r=5000\,{\rm km}$.}
\label{fig:IMHP}
\end{figure*}

The results of our multi-angle, $3 \times 3$ simulations in a particular case are shown in Fig.~\ref{fig:IMHP} for the inverted neutrino mass hierarchy.  We have chosen the values $\theta_{1 3} = 0.1$, $\theta_{1 2} = 0.6$, $\theta_{2 3} = \pi/4$, $\Delta m^{2}_{21} = 8.0 \times 10^{-5}\,{\rm eV}^{2}$, and $\Delta m^{2}_{32} = -3.0\times 10^{-3}\,{\rm eV}^{2}$ for this particular calculation to facilitate comparison with the choice of mixing parameters in Refs.~\cite{Duan08,Dasgupta:2008qy}.  

All of our simulations of the neutronization neutrino pulse from an ONeMg core-collapse supernova begin at an initial radius of $r = 900\,{\rm km}$, where the matter density is still large enough that no flavor transformation has yet taken place.  In order to compare our results directly with the results of Ref.~\cite{Duan08}, we use the density profile from the pre-collapse calculations in Refs.~\cite{Nomoto84,Nomoto87} in a post-bounce epoch.  However, the epoch in which we perform the calculations is $\sim 10\,{\rm ms}$ post-bounce, while the free-fall timescale at the radius $\left( 900\,{\rm km}\right)$ at which we begin our calculations is $\tau \sim 100\,{\rm ms}$, so there is not much time for non-homologous modifications to the density profile.  Moreover, the material at this radius is likely falling more slowly than the free fall rate.  In any case, the objective of our calculations is to compare single and multi-angle treatments of neutrino flavor transformation.  Ultimately, neutrino flavor transformation simulations must be performed in a consistent supernova model.

We model the neutrino emission as originating from a uniform sphere at a radius of $R_{\nu} = 60\,{\rm km}$ above the core of the proto-neutron star.  $R_{\nu}$ is the calculated radius of the \lq\lq neutrinosphere\rq\rq , where the electron neutrino optical depth is equal to unity.  It has been observed that the neutrinosphere radius is not a sharp surface, but instead is partially smeared out depending on neutrino energy as discussed in Ref.~\cite{Mezzacappa:1998qy}.  In the case of ONeMg core-collapse supernovae the matter density profile falls so swiftly with radius that this effect is small compared to the radius of the neutrinosphere itself and also the distance above the neutrino-sphere where significant flavor transformation takes place. % Although it is not present in these calculations, inclusion of this effect is planned for future versions of the BULB and FLAT codes.

In Subsection B, we present results of calculations with more recent and accurate values for neutrino mass-squared differences.  Figure~\ref{fig:IMHP} presents  the electron flavor neutrino survival probability $P_{\nu_{\rm e}\nu_{\rm e}}$, as a function of the cosine of the emission angle $\vartheta_{\rm R}$.  It also shows the vacuum mass basis angle-averaged neutrino energy spectra.  All of these results are at a radius $r = 5000\,{\rm km}$.  In the simulations used to generate this figure we have followed Ref.~\cite{Duan08}, and we have approximated the flavor content of the neutronization burst for the O-Ne-Mg core-collapse as pure $\nu_{\rm e}$.

Our simulations show that neutrino flavor evolution in the inverted mass hierarchy agrees  with what was found in Ref.~\cite{Duan08} in broad brush and also seems to agree with the explanation offered in Ref.~\cite{Dasgupta:2008qy}.  A clear step-wise swap occurs between the $\nu_{\rm 2}/\nu_{\rm 1}$ mass states at $E_{\nu} \approx 11\,{\rm MeV}$.  As with the single-angle results, there is no large-scale neutrino flavor transformation generated by mixing at the $\Delta m^{2}_{\rm atm}$ scale.  

\begin{figure}
\centering
\includegraphics[scale=.65]{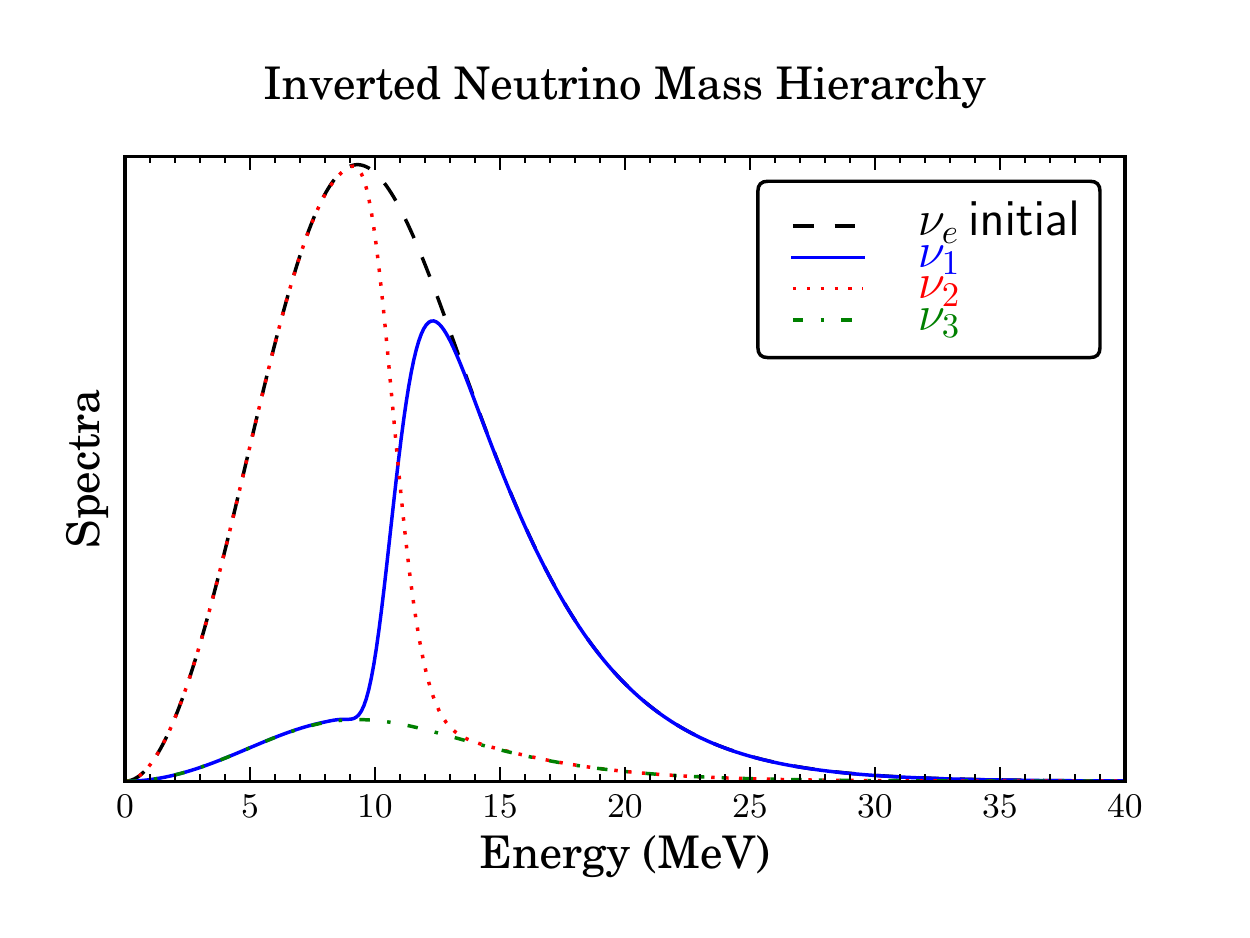}%scale = .63
\caption{Emission angle-averaged neutrino energy distribution functions versus neutrino energy plotted in the neutrino mass basis for the $3 \times 3$ multi-angle calculation of neutrino flavor evolution.  Results shown at a radius of $r = 5000 {\rm km}$.  In this simulation, a small $\left(10\%\right)$ admixture of all other species of neutrinos and anti-neutrinos are included.  }
\label{fig:IMHPadmix}
\end{figure}

It has been noted that in the inverted neutrino mass hierarchy the neutrino flavor field is unstable, in analogy to the way a pendulum balanced in the inverted position would be~\cite{Hannestad06,Duan07a}.  Flavor transformation in the other neutrino and anti-neutrino species is non-linearly coupled to mixing at the $\Delta m^{2}_{\rm atm}$ scale through the neutrino self-coupling potentials.  This coupling provides the impetus that drives the $\Delta m^{2}_{\rm atm}$ flavor isospins away from their unstable equilibrium.  However, the lack of any neutrino species besides $\nu_{\rm e}$ in the calculation shown in Fig.~\ref{fig:IMHP} prevents any perturbation of the \lq\lq inverted flavor pendulum,\rq\rq\ leaving it essentially balanced.  The inclusion of other neutrino and  anti-neutrino species, which have energy luminosities $L \sim 0.1 L_{\nu_{\rm e}}$ at this epoch, does not significantly affect these results and conclusions.  Figure~\ref{fig:IMHPadmix} shows that relatively small numbers ($\sim 10\%$ admixture) of these neutrinos are not capable of destabilizing the $\Delta m^{2}_{\rm atm}$ neutrino flavor mixing.  

\begin{figure*}
\centering
\includegraphics[scale=.85]{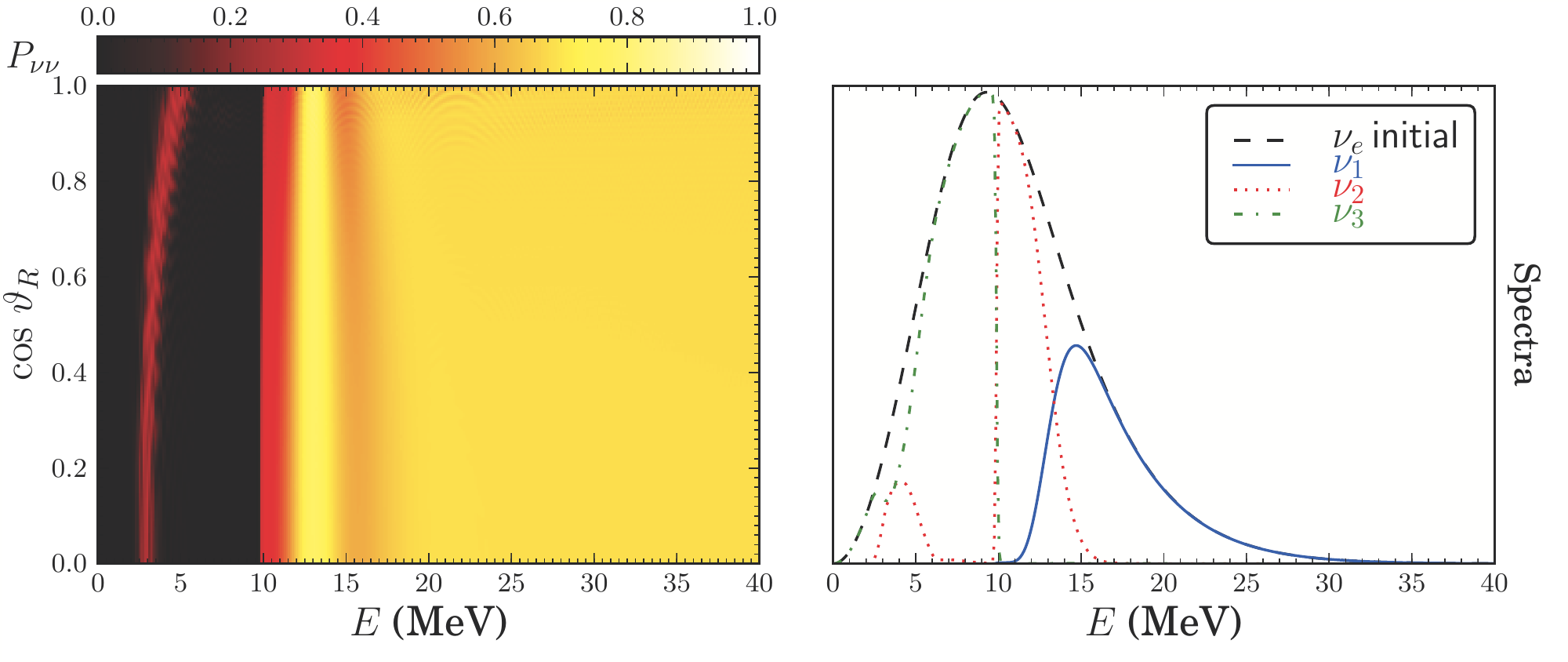}%scale = .63
\caption{Left panel: electron neutrino survival probability $P_{\nu_{\rm e}\nu_{\rm e}}$ (color/shading key at top left) for the normal mass hierarchy is shown as a function of cosine of emission angle, $\cos{\vartheta_{\rm R}}$, and neutrino energy, $E$ in MeV, plotted at a radius of $r=5000\,{\rm km}$.  Right:  mass basis (key top right, inset) emission angle-averaged neutrino energy distribution functions versus neutrino energy, $E$.  The dashed curve gives the initial $\nu_{\rm e}$ emission angle-averaged energy spectrum.  Movies of this simulation can be found at the URL in Ref.~\cite{Duan:web}.  Each frame of the movie shows a representation of the neutrino survival probability in various different bases at a fixed radius above the core.  Each successive frame is $1\,{\rm km}$ further out from the initial radius of $r_{\rm init} = 900\,{\rm km}$ out to the final radius $r=5000\,{\rm km}$.}
\label{fig:NMHP}
\end{figure*}

Figure~\ref{fig:NMHP} shows the results of the multi-angle simulations in the normal mass hierarchy.  We have used the same mixing parameters and initial conditions employed in the inverted neutrino mass hierarchy calculation, except that the sign of the atmospheric mass-squared difference has been reversed, $\Delta m^{2}_{32} = +3.0\times 10^{-3}\,{\rm eV}^{2}$.  Again, these calculations agree in broad brush with the results obtained for the single-angle simulations reported in Ref.~\cite{Duan08}.  In our multi-angle simulations the $\nu_{\rm 3}/\nu_{\rm 2}$ swap occurs at an energy of $E_{\nu} \approx 10\,{\rm MeV}$. This swap arises from the $\Delta m^{2}_{\rm atm}$ splitting.  Our simulations show that the second swap, corresponding to $\nu_{\rm 2}/\nu_{\rm 1}$, is at an energy of $E_{\nu} \approx 13.5\,{\rm MeV}$.  This swap arises from collective oscillations generated by mixing at the $\Delta m^{2}_{\odot}$ scale.  

In the case of the multi-angle calculation, both of the flavor swaps observed in the normal neutrino mass hierarchy occur at lower energies than their counterparts in the single-angle calculations.  Comparing the different simulations we find that for the $\nu_{\rm 3}/\nu_{\rm 2}$ swap $E^{\rm multi-angle}_{\rm swap} \approx 10.0\,{\rm MeV}$ and $E^{\rm single-angle}_{\rm swap} \approx 12.5\,{\rm MeV}$, and for the $\nu_{2}/\nu_{1}$ swap $E^{\rm multi-angle}_{\rm swap} \approx 13.5\,{\rm MeV}$ and $E^{\rm single-angle}_{\rm swap} \approx 15.0\,{\rm MeV}$.  

Interestingly, the inverted neutrino mass hierarchy does not exhibit any change in the observed $\nu_{2}/\nu_{1}$ swap energy between the multi-angle and the single-angle calculations.  The spectra we observe for the multi-angle calculations in the inverted neutrino mass hierarchy agree with the framework developed by Ref.~\cite{Dasgupta:2008qy}.  In that work, a single-angle toy model calculation showed that the resultant spectra of the neutronization neutrino burst for an O-Ne-Mg core-collapse supernova was formed by an initial phase of non-adiabatic, synchronous MSW flavor transformation, followed by collective neutrino oscillations that form the swaps observed in the simulations.  

The normal neutrino mass hierarchy case is quite different.  The final spectra in Fig.~\ref{fig:NMHP} show that there are $\sim 10\%$ fewer neutrinos that remain in mass state 3.  This implies that the flavor evolution in the multi-angle calculation was less adiabatic than that in the single-angle calculation.

%This picture of flavor transformation in the neutronization neutrino burst predicts that, if anything, a multi-angle calculation should experience a phase of synchronous MSW flavor transformation that is slightly more adiabatic than in the single-angle case.  The final spectra in Fig.~\ref{fig:NMHP} show that there are $\sim 10\%$ fewer neutrinos that remain in mass state 3, meaning that the flavor evolution of the multi-angle calculation was less adiabatic than that in the single-angle calculation.  These results are contrary to expectations, and are currently being investigated.

%\begin{figure}
%\centering
%\includegraphics[scale=.65]{./Spectraframe_massbasis0.eps}%scale = .63
%\caption{Emission angle-averaged neutrino energy distribution functions versus neutrino energy plotted in the neutrino mass basis for the $2 \times 2$ multi-angle calculation of neutrino flavor evolution in the $\Delta m^{2}_{\rm atm}$ mixing sector.  Results shown at a radius of $r = 5000 {\rm km}$.}
%\label{fig:2x2th12}
%\end{figure}

%\begin{figure}
%\centering
%\includegraphics[scale=.65]{./Spectraframe_massbasis1.eps}%scale = .63
%\caption{Emission angle-averagerd neutrino energy distribution functions versus neutrino energy plotted in the neutrino mass basis for the $2 \times 2$ multi-angle calculation of neutrino flavor evolution in the $\Delta m^{2}_{\odot}$ mixing sector.  Results shown at a radius of $r = 5000 {\rm km}$.}
%\label{fig:2x2th13}
%\end{figure} 

To illustrate the importance of using full $3 \times 3$ flavor mixing in the case of an O-Ne-Mg core-collapse supernova, we have performed a $2 \times 2$ multi-angle calculation for each mixing scale in the normal neutrino mass hierarchy.  Briefly, the swap energy of the $\Delta m^{2}_{\rm atm}$ mixing scale is $E_{\rm swap} \approx 10.0\,{\rm MeV}$ in our $2 \times 2$ multi-angle calculation, which corresponds closely with the swap energy found in the $3 \times 3$  multi-angle flavor mixing case.  However, the swap energy of the $\Delta m^{2}_{\odot}$ mixing scale is found to be $E_{\rm swap} \approx 11.0\,{\rm MeV}$ for the $2 \times 2$ multi-angle calculation, which is lower than the corresponding  $\nu_{2}/\nu_{1}$ swap $E_{\rm swap} \approx 13.5\,{\rm MeV}$ of the $3 \times 3$ multi-angle calculation.  

\subsection{Sensitivity to Neutrino Mass-Squared Differences}

\begin{figure*}
\centering
\includegraphics[scale=.85]{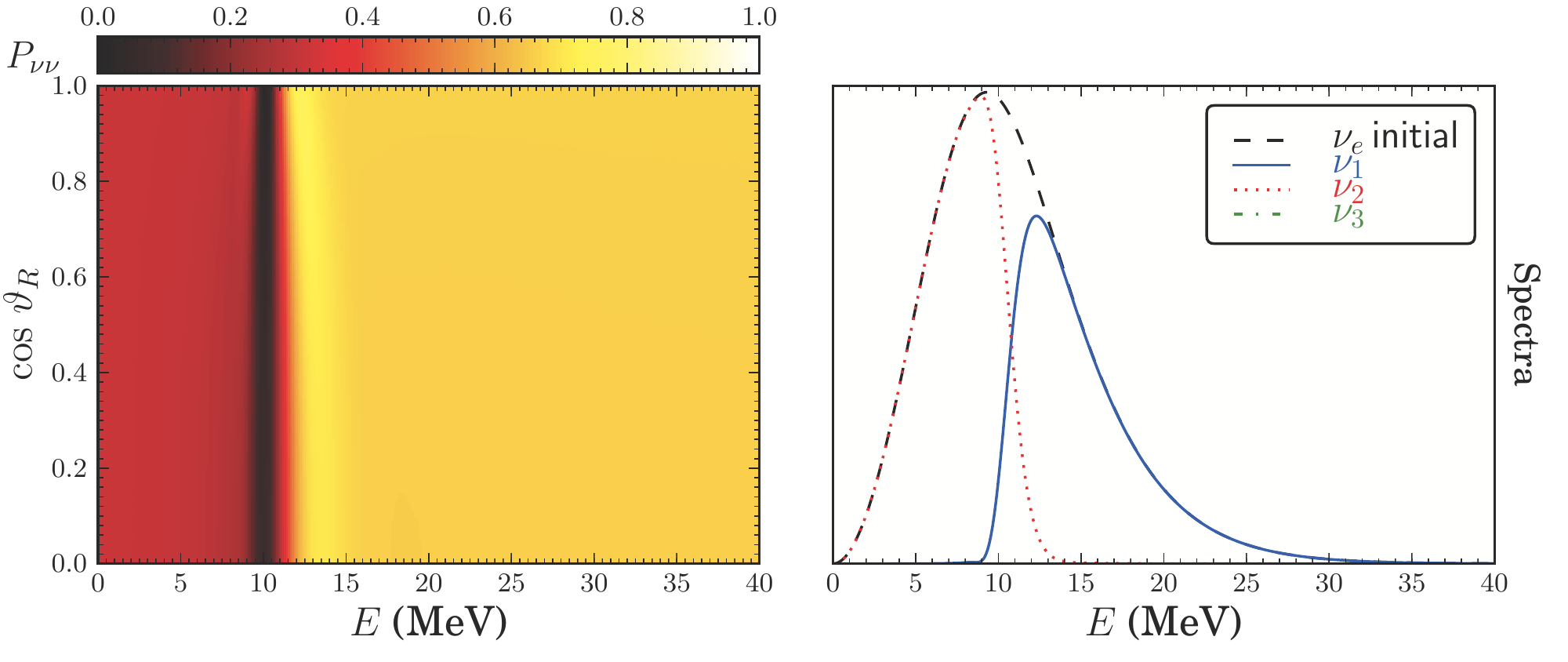}%scale = .63
\caption{This calculation of the flavor evolution of neutrinos in the inverted mass hierarchy was conducted with values of $\Delta m^{2}_{\odot} = 7.6 \times 10^{-5}\,{\rm eV}^{2}$ and $\Delta m^{2}_{\rm atm} = -2.4\times 10^{-3}\,{\rm eV}^{2}$.  Left panel: electron neutrino survival probability $P_{\nu_{\rm e}\nu_{\rm e}}$ (color/shading key at top left) for the inverted mass hierarchy is shown as a function of cosine of emission angle, $\cos{\vartheta_{\rm R}}$, and neutrino energy, $E$ in MeV, plotted at a radius of $r=5000\,{\rm km}$.  Right:  mass basis (key top right, inset) emission angle-averaged neutrino energy distribution functions versus neutrino energy, $E$.  The dashed curve gives the initial $\nu_{\rm e}$ emission angle-averaged energy spectrum.}
\label{fig:NV_IMHP}
\end{figure*}

Since the publication of the original set of papers on the neutronization neutrino burst of an O-Ne-Mg core-collapse supernova~\cite{Duan08,Dasgupta:2008qy}, the experimental limits on $\Delta m^{2}_{\odot}$ and $\Delta m^{2}_{\rm atm}$ have been refined, resting currently at the values of $\Delta m^{2}_{\odot} = 7.6 \times 10^{-5}\,{\rm eV}^{2}$ and $\Delta m^{2}_{\rm atm} = 2.4\times 10^{-3}\,{\rm eV}^{2}$~\cite{Maltoni:2007lr}.  We have conducted a set of $3 \times 3$ multi-angle calculations to ascertain the effect the new experimental constraints would have on the flavor evolution history of the neutrinos released in the neutronization burst.  Figure~\ref{fig:NV_IMHP} shows the results of our calculations for neutrino flavor transformation in the inverted neutrino mass hierarchy.  There is no significant change observed in the final state of neutrinos at a radius of $r = 5000\,{\rm km}$ relative to our earlier calculations with $\Delta m^{2}_{21} = 8.0 \times 10^{-5}\,{\rm eV}^{2}$.

\begin{figure*}
\centering
\includegraphics[scale=.85]{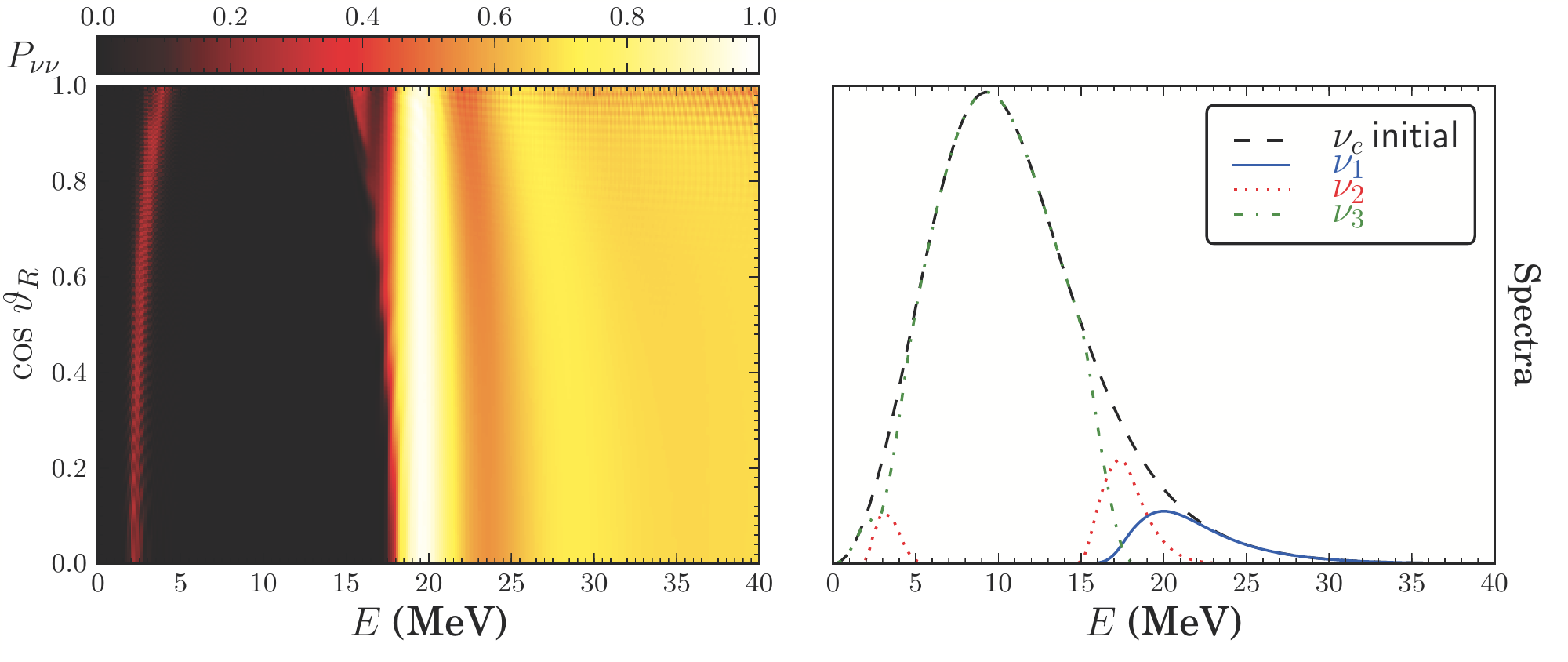}%scale = .63
\caption{This calculation of the flavor evolution of neutrinos in the normal mass hierarchy was conducted with values of $\Delta m^{2}_{\odot} = 7.6 \times 10^{-5}\,{\rm eV}^{2}$ and $\Delta m^{2}_{\rm atm} = 2.4\times 10^{-3}\,{\rm eV}^{2}$.  Left panel: electron neutrino survival probability $P_{\nu_{\rm e}\nu_{\rm e}}$ (color/shading key at top left) for the normal mass hierarchy is shown as a function of cosine of emission angle, $\cos{\vartheta_{\rm R}}$, and neutrino energy, $E$ in MeV, plotted at a radius of $r=5000\,{\rm km}$.  Right:  mass basis (key top right, inset) emission angle-averaged neutrino energy distribution functions versus neutrino energy, $E$.  The dashed curve gives the initial $\nu_{\rm e}$ emission angle-averaged energy spectrum.}
\label{fig:NV_NMHP}
\end{figure*}

Figure~\ref{fig:NV_NMHP} shows the results of our calculations for neutrino flavor transformation in the normal neutrino mass hierarchy with the latest mass-squared values.  With the new neutrino mixing parameters there is an increase in the observed swap energies for mixing at both the $\Delta m^{2}_{\odot}$ and $\Delta m^{2}_{\rm atm}$ scales over the cases with the original mixing parameters adopted in~\cite{Duan08}.  The $\nu_{\rm 3}/\nu_{\rm 2}$ swap occurs at an energy of $E_{\nu} \approx 16.5\,{\rm MeV}$, while the $\nu_{\rm 2}/\nu_{\rm 1}$ swap  is pushed even higher to an energy of $E_{\nu} \approx 19.0\,{\rm MeV}$, as opposed to $E_{\nu} \approx 10.0\,{\rm MeV}$ and $E_{\nu} \approx 13.5\,{\rm MeV}$ respectively. 

\begin{figure}
\centering
\includegraphics[scale=.45]{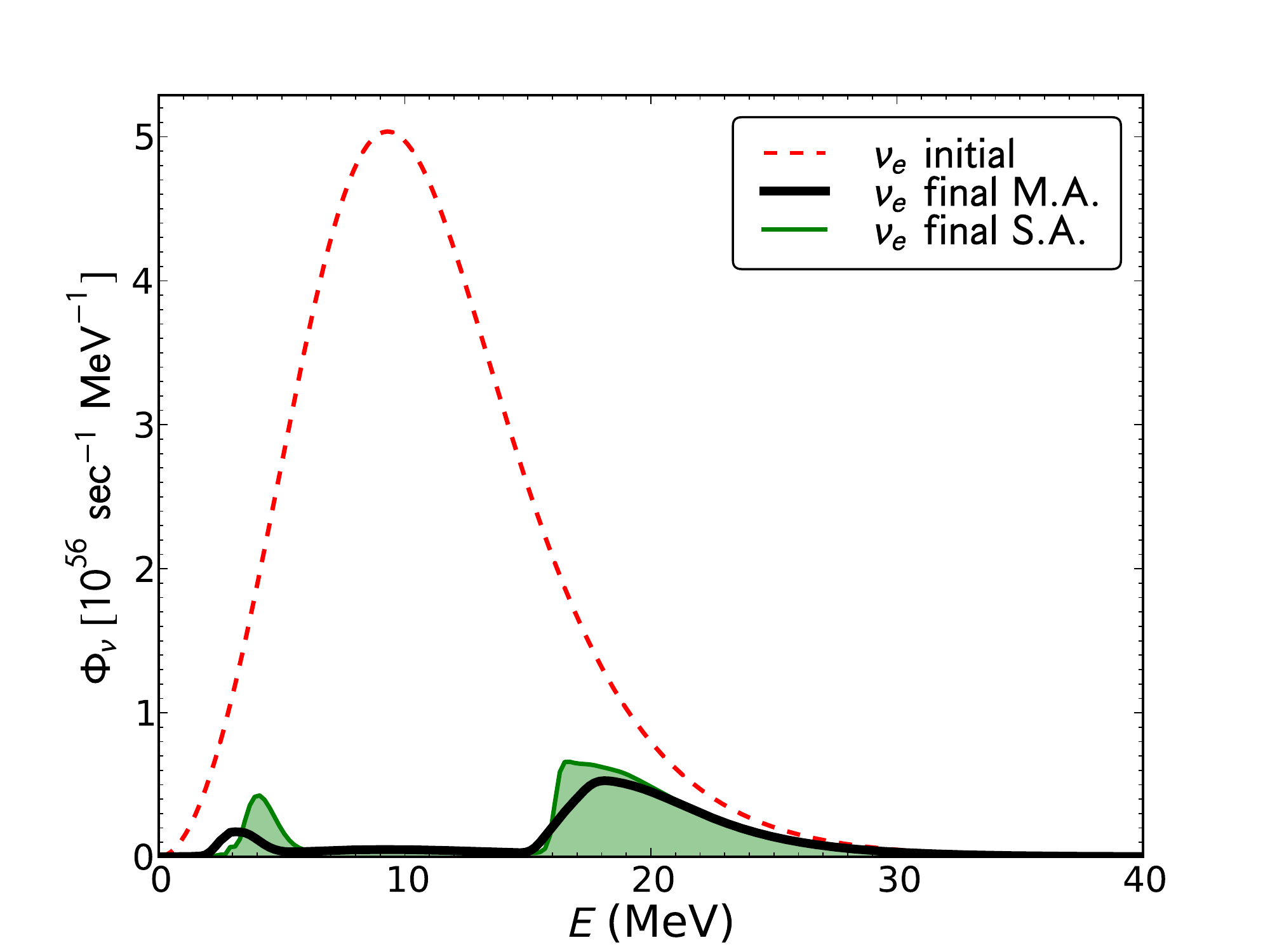}%scale = .63
\caption{Emission angle-averaged electron neutrino flux $\Phi_{\nu}$ (key top right, inset) for the normal neutrino mass hierarchy is shown as a function of neutrino energy $E$ in MeV.  The dashed curve gives the initial $\nu_{\rm e}$ emission angle-averaged neutrino flux.  The shaded region gives the predicted flux in a single-angle calculation, and the thick line shows the flux predicted by the multi-angle calculation.  These calculations of electron neutrino flux are done using $\Delta m^{2}_{\odot} = 7.6 \times 10^{-5}\,{\rm eV}^{2}$ and $\Delta m^{2}_{\rm atm} = -2.4\times 10^{-3}\,{\rm eV}^{2}$ and $\theta_{13} = 0.1$.}
\label{fig:flux_nh}
\end{figure}

\begin{figure}
\centering
\includegraphics[scale=.45]{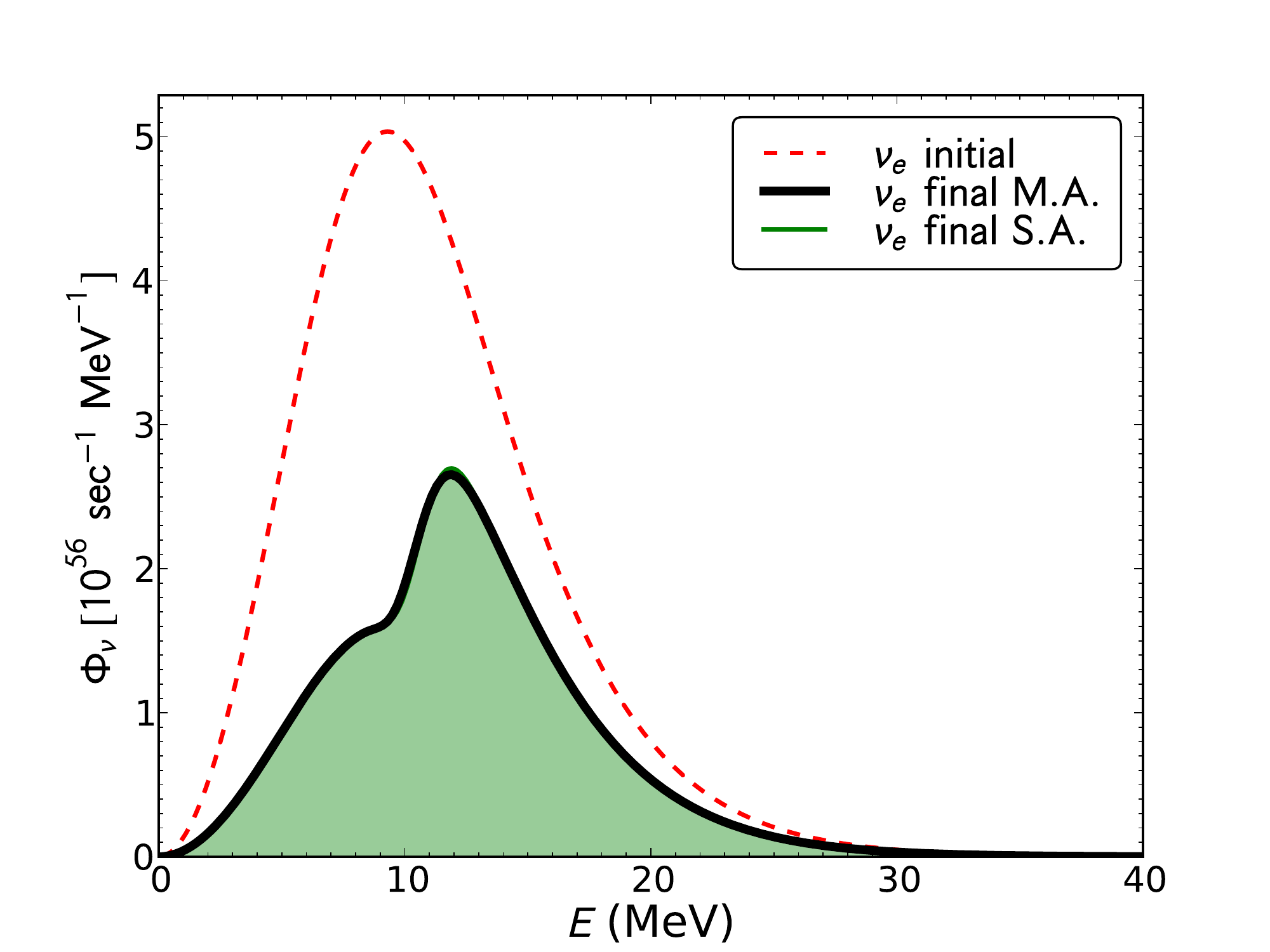}%scale = .63
\caption{Emission angle-averaged electron neutrino flux $\Phi_{\nu}$ (key top right, inset) for the inverted neutrino mass hierarchy is shown as a function of neutrino energy $E$ in MeV.  The dashed curve gives the initial $\nu_{\rm e}$ emission angle-averaged neutrino flux.  The shaded region gives the predicted flux in a single-angle calculation, and the thick line shows the flux predicted by the multi-angle calculation.  These calculations of electron neutrino flux are done using $\Delta m^{2}_{\odot} = 7.6 \times 10^{-5}\,{\rm eV}^{2}$ and $\Delta m^{2}_{\rm atm} = -2.4\times 10^{-3}\,{\rm eV}^{2}$ and $\theta_{13} = 0.1$.}
\label{fig:flux_ih}
\end{figure}

Because detection and characterization of supernova neutrinos is the ultimate aim of research in this field, we have included Fig.~\ref{fig:flux_nh} and Fig.~\ref{fig:flux_ih} to show predictions for the flux of electron neutrinos emitted by the neutronization neutrino burst of an O-Ne-Mg core collapse supernova.  Figure~\ref{fig:flux_nh} shows the predictions of both the single-angle and multi-angle calculations for the normal neutrino mass hierarchy, while Fig.~\ref{fig:flux_ih} shows the same for the inverted neutrino mass hierarchy.

\subsection{Variation of $\theta_{13}$}

\begin{figure}
\centering
\includegraphics[scale=.73]{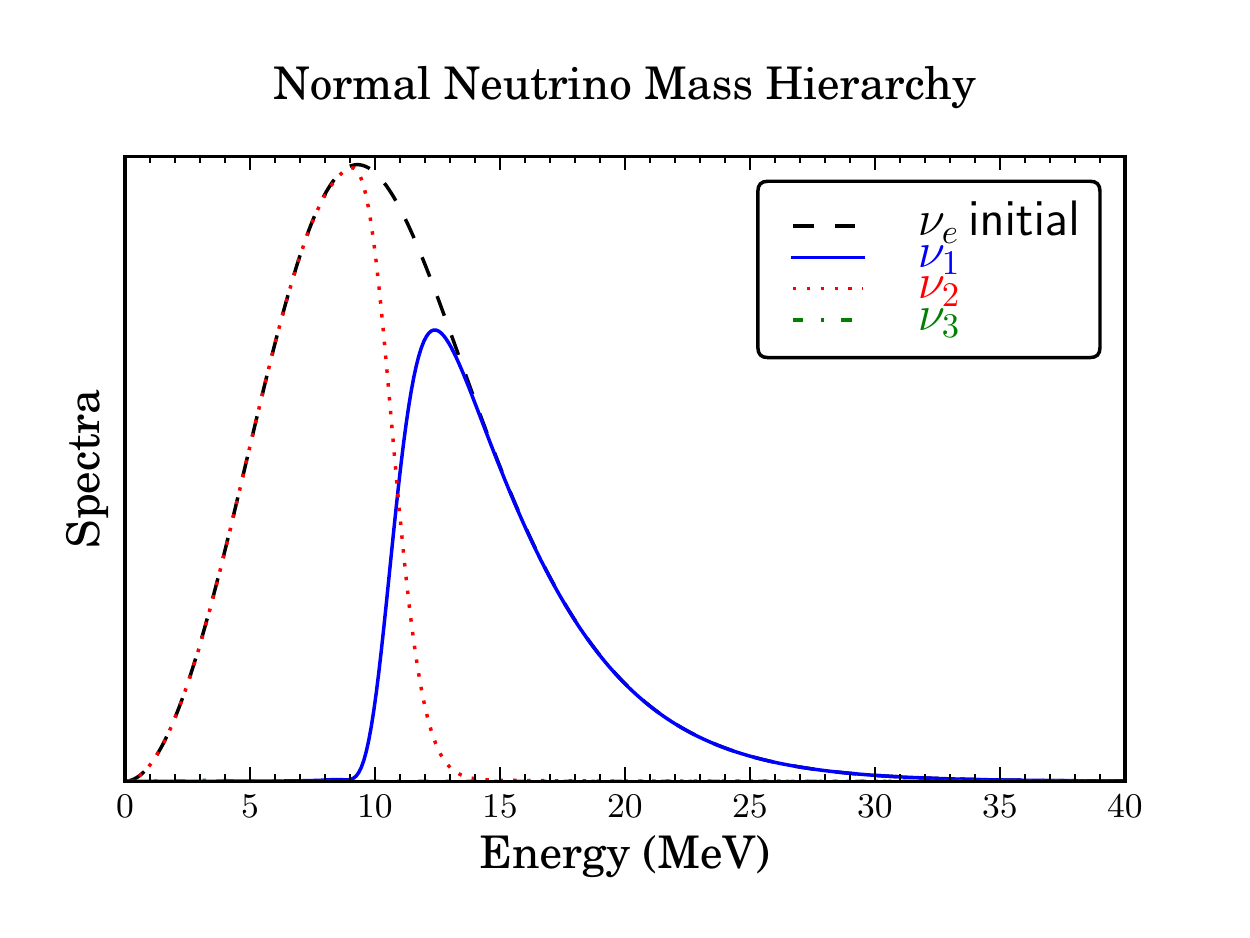}%scale = .63
\caption{Emission angle-averaged neutrino energy distribution functions versus neutrino energy plotted in the neutrino mass basis for the $3 \times 3$ multi-angle calculation of neutrino flavor evolution.  This calculation employs a significantly reduced value of $\theta_{13} = 1.0 \times 10^{-3}$.  This mixing angle is associated with flavor transformation at the $\Delta m^{2}_{\rm atm}$ scale.  Results shown at a radius of $r = 5000 {\rm km}$.}
\label{fig:lowth13}
\end{figure}

Of particular interest is the effect produced by a significantly decreased $\theta_{13}$ mixing angle.  As has been found in previous work, reducing the value of $\theta_{13}$ can affect sensitively the energy of the flavor swap at the $\Delta m^{2}_{\rm atm}$ scale for the normal neutrino mass hierarchy~\cite{Duan07c}.  We performed a new calculation that matched the initial conditions of our primary, $\nu_{\rm e}$ only, simulation of the neutronization burst, but this time with a value of $\theta_{13} = 1.0 \times 10^{-3}$, two orders of magnitude smaller than the $\theta_{13}$ value in the original calculation.  This new value of $\theta_{13}$ drives the neutrino background-enhanced flavor evolution at the $\Delta m^{2}_{\rm atm}$ scale entirely non-adiabatic, and Fig.~\ref{fig:lowth13} shows that this produces a resultant neutrino energy spectrum that is essentially indistinguishable from the spectrum produced in our calculations for the inverted neutrino mass hierarchy.  This may present complications in the future determination of the neutrino mass hierarchy and $\theta_{13}$ using the neutronization pulse of an O-Ne-Mg core-collapse supernova.  Planned earth-based experiments should be able to set an upper limit on the value of $\sin^{2}2\theta_{13} \sim 1.0 \times 10^{-2}$~\cite{Daya-Bay:2007uq}.  The neutronization neutrino pulse signal from an O-Ne-Mg core-collapse supernova may have limited ability to discern a value of $\theta_{13}$ much below this threshold.  On the other hand, the late-time supernova neutrino signal, which should be more or less generic for all core-collapse supernovae, has no such limitations.  Detection of the sense of the swap in this case should provide an unambiguous determination of the neutrino mass hierarchy.  Measurements like these could be complementary to laboratory based neutrino mass probes.  For example, experiments such as the MAJORANA project may be able to determine the neutrino mass hierarchy directly~\cite{Elliott:2009fj}.

\section{Conclusion}
We have reported the first results of a fully $3 \times 3$, multi-angle simulation of the flavor evolution of neutrinos emitted from a supernova event.  This case study examined the neutrino flavor oscillations in the neutronization pulse emitted from an O-Ne-Mg core-collapse supernova.  In these events, the matter density above the core has such a steep gradient that both neutrino self-coupling and the overlap of neutrino MSW resonance regions must be accounted for to obtain accurate energy spectra of the neutrino signals.  

We have developed a suite of programs that are capable of computing the solution to the non-linear, Schr\"odinger-like equation of motion that governs the neutrino flavor evolution in the coherent regime.  As a result of the nature of the O-Ne-Mg core-collapse supernova environment, our codes have been pushed to a new level of complexity, handling the evolution of all three neutrino flavors with the inclusion of a full treatment of the angular dependence of both the neutrino-neutrino neutral current forward exchange scattering and neutrino flavor evolution history.  By approaching the solution to this problem with two independent sets of solution algorithms (the BULB and FLAT codes), we have verified that the results of our calculations are entirely self-consistent.

The features of the neutrino flavor transformation history for the neutronization neutrino burst observed in our calculations are partially consistent with what has been found in single-angle calculations.  In the normal neutrino mass hierarchy, for $\theta_{13} = 0.1$, a pair of flavor swaps are observed, corresponding to the $\Delta m^{2}_{\odot}$ and $\Delta m^{2}_{\rm atm}$ mixing scales.  In contrast, for this $\theta_{13}$, the  inverted neutrino mass hierarchy shows only a single flavor swap, originating from collective neutrino oscillations at the $\Delta m^{2}_{\odot}$ scale.  Our studies show that when $\theta_{13}$ is reduced to the value of $\theta_{13} = 0.001$, the double swap observed in the normal neutrino mass hierarchy becomes a single flavor swap generated by mixing at the $\Delta m^{2}_{\odot}$ scale, identical to the observed neutrino flavor energy spectrum calculated for the inverted neutrino mass hierarchy.

In the normal neutrino mass hierarchy in particular, it is critical that $3 \times 3$ flavor evolution be employed.  Moreover, we have also shown that our results for the neutrino flavor swap energies retain their relevance throughout the evolution of the neutronization neutrino burst itself, as the added luminosity of neutrino species besides $\nu_{\rm e}$, which rise late in the duration of the burst, do not significantly affect the energy of the swaps.

We have shown that the energy of the flavor swaps can be sensitive to the values of $\Delta m^{2}$ at the $\sim 10\%$ level.  This gives added importance to continuing laboratory efforts to measure $\Delta m^{2}$.  These constraints will only improve with time.

It has been a hope that the collective neutrino oscillation signatures, if detected, could reveal the neutrino mass hierarchy and/or $\theta_{13}$.   However, discerning the neutrino mass hierarchy from an O-Ne-Mg core-collapse neutronization burst signal would depend on distinguishing two spectral swaps from one spectral swap.  Moreover, this would have to be done at relatively low neutrino energies, and likely also require both charged and neutral current detection capabilities.  It is not clear that all proposed future supernova neutrino detectors would be capable of this.  Our simulations also show that the normal mass hierarchy will give rise to only a single swap if $\theta_{13}$ is sufficiently small.  This could make resolving the hierarchy problematic in a detected O-Ne-Mg core-collapse neutronization burst, depending on the value of $\theta_{13}$.  However, this limitation is likely not present in the late time neutrino signals of any core-collapse event. 

From the unexpected behavior of the swap energies in the normal neutrino mass hierarchy, it is clear that (as has been seen many times before in computational physics) incorporation of more realistic description of the physical environment in our simulations has resulted in new phenomenology.  There is an intimate interplay at work between the competing effects of neutrino self-coupling and matter driven flavor evolution taking place in the heart of every exploding star.  While the single-angle approximation remains a well understood and useful tool in evaluating the flavor evolution of neutrinos emitted from supernovae with relatively dense envelopes, those types of events may not be as common as previously believed.  Given recent advances in the understanding of the hydrodynamics that governs core-collapse supernovae, it is clear that any hope of understanding the neutrino signal from stars with thin envelopes may require even more detailed treatment~\cite{Burrows:2009fk,Bruenn:2009uq,Bruenn:2010qy,Hammer:2010yq,Muller:2010kx,Ott:2009fj}.

 In a sense our work is a step backward.  At least for the particular supernova model we consider here, we must conclude that single-angle treatments are inadequate when neutrinos have the normal mass hierarchy.  It is still mysterious why the single-angle approach is efficacious in some regimes and inadequate in others.  In any case, the work presented here may be a first step towards understanding this issue.

\section{Acknowledgments}
We would like to thank A. Friedland, C. Lunardini, G. Raffelt, and S. Skory for valuable conversations.  This work was supported in part by NSF grant PHY-06-53626 at UCSD, DOE grant DE-FG02-87ER40328 at the UMN, and by the DOE Office of Nuclear Physics, the LDRD Program and Open Supercomputing at LANL, and an Institute of Geophysics and Planetary Physics/LANL minigrant.  In addition the research of H.D. is supported by the LANL Laboratory Dierected Research and Development program through the director's postdoctoral fellowship at LANL.

\bibliography{onemg}

\end{document}